# Iron melting curve with a tricritical point

A. AITTA

*Institute of Theoretical Geophysics, Department of Applied Mathematics and Theoretical Physics, University of Cambridge, Wilberforce Road, Cambridge CB3 0WA, UK*

e-mail: A.Aitta@damtp.cam.ac.uk

Solidification as a first order phase transition is described in the Landau theory by the same equation as tricritical phenomena. Here, the solidification or melting temperature against pressure curve is modelled to end at a tricritical point. The model gives the phase transition temperature's dependence on pressure up to the quadratic term with a definite expression for the coefficients. This formula is expected to be generally valid for pure materials having melting curves with $dT/dP$ approaching zero at very high $P$. Excellent experimental agreement is obtained for iron, the material having the most high pressure data which rather accurately determines the value of the coefficient defining the curvature. The geophysically interesting iron solidification temperatures at the Earth's core pressures are obtained. In addition, the general formulae for entropy change, latent heat and volume contraction in solidification are found and calculated for iron as functions of pressure and temperature.

## 1. Introduction

The states of a one-component substance such as iron are schematically presented in figure 1 in a temperature against pressure diagram. The vaporization curve where the vapour phase and the liquid phase coexist is known to end at high temperatures at a critical point ($P_c$, $T_c$) beyond which distinct vapour and liquid phases do not exist. The fluid beyond temperature $T_c$ is called supercritical: for instance, it is not possible to liquefy it by increasing the pressure, only by cooling. The $T_c$ boundary of the supercritical fluid is indicated by the dashed line as in [1]. Here, in addition, it is boldly continued up to and past the pressures of the melting/solidification curve, where the liquid phase and the solid phase coexist. In this paper, it is assumed that the phase transition curve ends in a tricritical point. Evidence consistent with this assumption is discussed. This work demonstrates that for iron we have currently enough experimental evidence to construct using Landau theory the melting curve with a tricritical end point, but at a temperature and pressure beyond present experimental techniques. The Landau theory determines important thermodynamic parameters along the melting curve, and gives predictions for the temperature and latent heat at pressures that occur in the Earth's core.

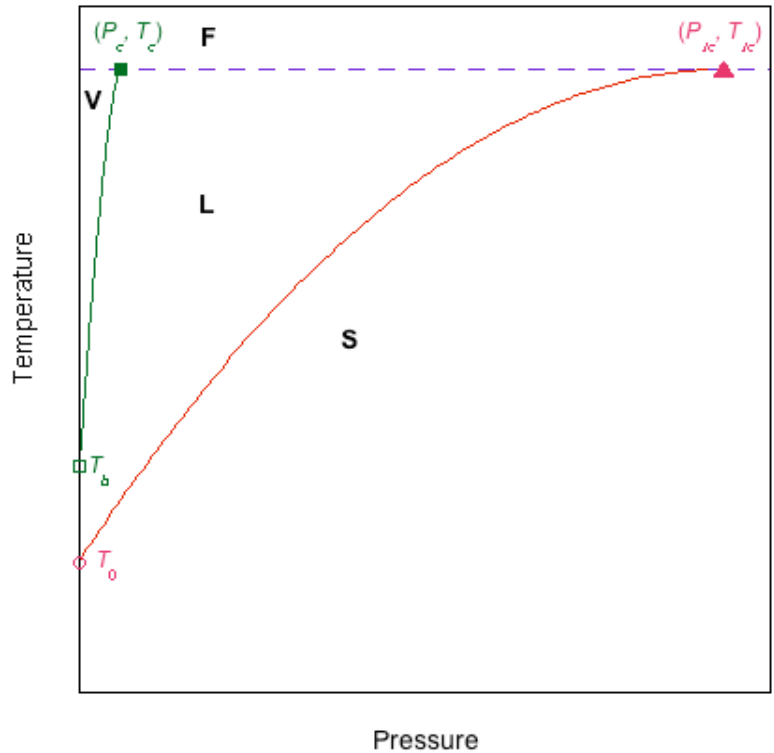

**Figure 1.** A schematic phase diagram for a material for whose melting curve *dT/dP* = 0 at high *P*. S is for solid phase, L for liquid phase, V for vapour phase and F for supercritical fluid phase. The vaporization curve ends at the critical point ($P_c$, $T_c$). The melting curve is considered here to end at the tricritical point ($P_{tc}$, $T_{tc}$). $T_0$ and $T_b$ are the zero pressure melting and boiling temperatures. $P_c$ is overestimated to make the vapour phase area visible when linear axes are used.

The possibility that there is a critical point also at the end of the melting curve has been of concern for over a century. Historically it was expected to exist by J. H. Poynting, M. Planck, W. Ostwald, P. P. Weimarn and J. J. van Laar [2]. However, the experimentally researched pressure range has not reached it for any three-dimensional material. Bridgman [2] noticed that the latent heat and volume difference between liquid and solid did not seem to decrease to zero at the same pressure as they should if there were a critical point. The latent heat stayed rather constant or even increased while the volume difference between liquid and solid decreased slightly in the pressure range then available. Bridgman did not appreciate that the latent heat could have a maximum at some pressure but could still decrease to zero at a critical point at much higher pressure. Instead, he first [2] claimed the existence of a critical end point was a remote possibility and later denied its existence altogether [3]. He tried to establish that the temperature along the melting curve would keep increasing with increasing pressure without having a critical point, maximum or an asymptotic value. In contrast, a tricritical point was predicted by the Lennard-Jones and Devonshire model [4]. In addition, two-dimensional matter such as xenon on graphite [5] demonstrates that the melting transition can exhibit a tricritical point. Also, some recent works [6-7] have challenged the shape of the melting curve: at least for some metals the curve seems to already have *dT/dP* approaching zero at pressures less than 100 GPa. Nevertheless, those pressures are still substantially less than what is available by means of the shock-wave research, and it would be important to establish whether *dT/dP* continues to approach zero at these higher pressures.

Iron is the material for which we have most high pressure data, mainly because of its role as the most important element in the Earth's core. However, that data has a rather high scatter. A consistent selection of the wide-spread iron data accumulated over the last 20 years has here been chosen to demonstrate that a tricritical point can exist at the end of the melting curve with *dT/dP* = 0. Landau theory gives a general expression for the solidification or melting temperature. It depends quadratically on pressure up to the tricritical pressure $P_{tc}$ where the curve changes to a second order transition line $T = T_{tc}$ where $T_{tc}$ is the temperature at the tricritical point. Presently, only iron seems to have good enough data in a wide enough pressure range to allow one to find the curvature of the melting curve accurately enough to determine the location of the tricritical point.

In addition, the change of entropy, latent heat and volume contraction in the solidification have been calculated. Using the experimentally known zero pressure values for the iron melting temperature and the latent heat, they are found at all pressures, including those at the Earth's core. These results are important in modelling the history and the present state of the Earth's core. The latent heat depends cubically on pressure and reaches a maximum at *P* about 280 GPa. Thus at small pressures the latent heat increases but it still decreases to zero at the tricritical pressure, together with the volume contraction.

## 2. Solidification using Landau theory

The tricritical point as a phenomenon was described mathematically in 1937 by Landau [8]. It has a special interest for physicists because by means of renormalization group theory its description has been shown to be exact in three spatial dimensions up to thermal fluctuations (see, for instance [9, 10]). At a tricritical point, the first order transition changes smoothly to be a second order transition while the sign of a coefficient in the Landau potential passes through zero. Here the structure of the coefficients of the Landau potential is constructed for a concave melting curve whose slope *dT/dP* approaches zero at high *P*.

The state of the system which depends on temperature *T* and pressure *P* is described by the Gibbs free energy $G = H - TS$ using the enthalpy *H* and the entropy *S*. Using this, the thermodynamic foundation of solidification is well established. Firstly, at the phase transition, *G* is the same for both liquid (subscript *l*) and solid (subscript *s*). Thus $H_l - TS_l = H_s - TS_s$. The latent heat or the heat of fusion in the transition is $L = H_l - H_s = T(S_l - S_s)$ which gives, using *ΔS* for the change in the entropy,

$$L = T\Delta S. \tag{1}$$

The total differential of the Gibbs free energy is $dG = -SdT + VdP$. Thus the entropy is

$$S = -\left(\frac{\partial G}{\partial T}\right)_P \tag{2}$$

and

$$V = \left(\frac{\partial G}{\partial P}\right)_T. \tag{3}$$

Secondly, along the phase coexistence curve, the change in the Gibbs free energy is equal for both phases: $-S_l dT + V_l dP = -S_s dT + V_s dP$. This can be reorganized to solve the temperature's pressure gradient using the changes of the volume and entropy in the phase transition:

$$\frac{dT}{dP} = \frac{\Delta V}{\Delta S}. \tag{4}$$

The inverse of this equation, with the substitution of $\Delta S$ from (1), is known as the Clausius-Clapeyron equation.

Thirdly, according to Landau [8] the Gibbs free energy density can be written as a dimensionless potential having an even power series

$$\Phi = \frac{1}{6}x^6 + \frac{1}{4}g x^4 + \frac{1}{2}\varepsilon x^2 + \Phi_0 \tag{5}$$

where $x$ is an order parameter ($= 0$ for liquid, $\neq 0$ for solid). For the first order phase transition, terms up to the sixth order are needed. Any higher order terms can be eliminated using coordinate transformations as in bifurcation theory [11]. $\Phi_0$ is the part of the Gibbs free energy independent of the order parameter; $\Phi_0$ as well as the coefficients $g$ and $\varepsilon$ generally depend on the physical parameters: on temperature and pressure for this solidification problem. Near the critical point, $\varepsilon$ depends linearly on temperature and $g$ linearly on temperature or pressure [12]. The equation (5) describes also the change from the first order to the second order transition when $g$ changes from negative to positive. The special situation where $g = \varepsilon = 0$ is called a tricritical point and is the organizing centre of the first order phase transitions.

In equilibrium, the order parameter takes a value where the potential has a local or global minimum. The minima of $\Phi$, the solutions of

$$x^5 + g x^3 + \varepsilon x = 0 \tag{6}$$

at which $\frac{d^2\Phi}{dx^2}$ is positive, give three stable states provided $0 < \varepsilon < \frac{1}{4}g^2$ and $g < 0$. They are at $x = 0$ and $x = \pm\sqrt{-\frac{g}{2} + \sqrt{\frac{g^2}{4} - \varepsilon}}$. The thermodynamic transition from liquid to solid occurs when all three minima of the potential are equally deep: $\Phi - \Phi_0 = 0$ at those three values of $x$. This requires

$$\varepsilon = \frac{3}{16}g^2 \tag{7}$$

or equivalently

$$g = -4\sqrt{\frac{\varepsilon}{3}} \tag{8}$$

and the minima are then at $x = 0$ and

$$x = \pm\sqrt{\frac{-3g}{4}} \tag{9}$$

or equivalently

$$x = \pm(3\varepsilon)^{1/4} \tag{10}$$

showing that the tricritical critical exponent is $\frac{1}{4}$. The $(\varepsilon, g)$ stability curve given by equation (7) is shown in figure 2. The liquid phase is then in thermal equilibrium with the solid phase. If $\varepsilon > \frac{3}{16}g^2$ the liquid is preferred, and if $\varepsilon < \frac{3}{16}g^2$ the solid.

The stability curve has at the tricritical point a tangent along the $g$–axis. Thus the materials with their melting curve tangent satisfying $dP/dT = 0$ at $P_{tc}$ have their $g$–axis parallel to the $P$–axis. Therefore one can define

$$g = \frac{P}{P_{tc}} - 1. \tag{11}$$

By taking the $\varepsilon$–axis perpendicular to the $g$–axis, $\varepsilon$ depends only on temperature:

$$\varepsilon = \frac{3}{16}\frac{T_{tc} - T}{T_{tc} - T_0} \tag{12}$$

where $T_0$ is the zero pressure melting temperature which is experimentally well-known.

The stability equation (7) gives the transition temperature from liquid to solid to be

$$T = T_0 + 2\frac{T_{tc} - T_0}{P_{tc}}P - \frac{T_{tc} - T_0}{P_{tc}^2}P^2, \text{ for } P \leq P_{tc}. \tag{13}$$

The numerical values for $T_{tc}$ and $P_{tc}$ may be obtained using experimental results to find the curvature of the melting curve. Especially, using the most reliable selection from the widespread iron data available at the moment one can determine $T_{tc}$ and $P_{tc}$ with an adequate accuracy.

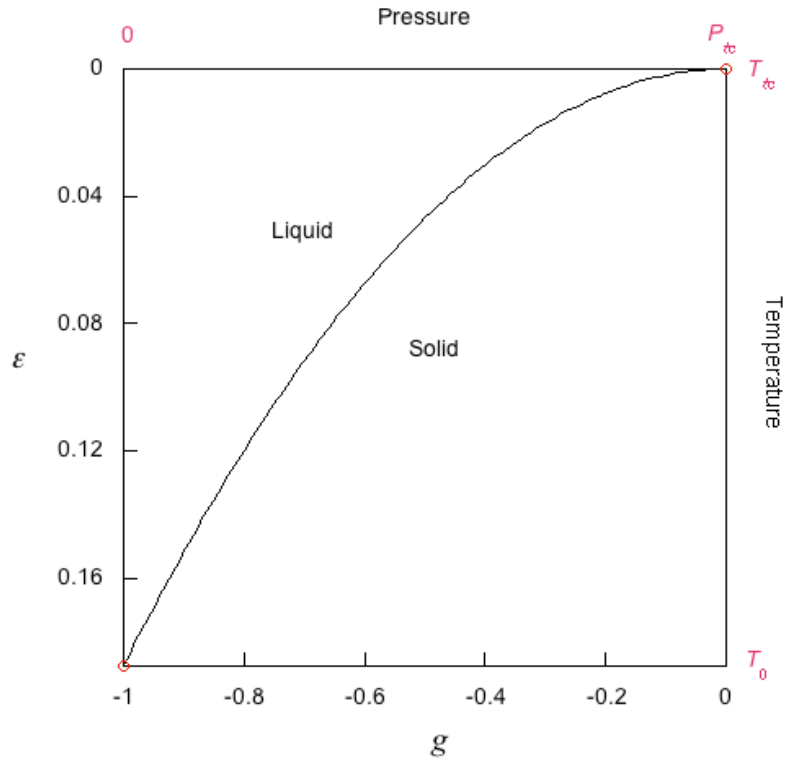

**Figure 2.** A generic liquid-solid phase diagram. Below the stability curve, $g = -4\sqrt{\frac{\varepsilon}{3}}$, the solid phase is stable. $g = \varepsilon = 0$ at the tricritical point ($P_{tc}$, $T_{tc}$). $T_0$ is the zero pressure melting temperature.

The expression (13) is similar to $T - T_0 = a\,P + b\,P^2$ found by Tammann to represent his experimental results as cited on page 196 in [2]. However, here the structure of the coefficients is also obtained, although the validity of the expression is limited to materials for which $dT/dP = 0$ at high $P$.

Other thermodynamic quantities can be found, too. In order to calculate the entropy from equation (2) one can use as $G$ the Landau potential $\Phi$ from (5) multiplied by a dimensional constant $\theta$ to gain comparability to the experimental results. First, one obtains $S = \theta \dfrac{\partial \Phi_0}{\partial T} - \tfrac{1}{2}\theta x^2 \dfrac{d\varepsilon}{dT}$ and from (12) $\dfrac{d\varepsilon}{dT} = \dfrac{-3}{16(T_{tc} - T_0)}$. Together these give $S = \theta \dfrac{\partial \Phi_0}{\partial T} + \dfrac{3}{32}\theta x^2 \dfrac{1}{T_{tc} - T_0}$. Since $x = 0$ for the liquid phase, $S_l = \theta \dfrac{\partial \Phi_0}{\partial T}$. For the solid phase one obtains from (10) and (12) that $x^2 = \dfrac{3}{4}\sqrt{\dfrac{T_{tc} - T}{T_{tc} - T_0}}$, giving for the change of entropy

$$\Delta S = \frac{9}{128} \frac{\theta}{T_{tc} - T_0} \sqrt{\frac{T_{tc} - T}{T_{tc} - T_0}} \tag{14}$$

or equivalently

$$\Delta S = \frac{9}{128} \frac{\theta}{T_{tc} - T_0} \frac{P_{tc} - P}{P_{tc}}. \tag{15}$$

Hence $\Delta S$ decreases linearly with increasing $P$ from $9\theta/[128(T_{tc} - T_0)]$ at $P = 0$ to 0 at $P = P_{tc}$.

From (13) one finds $\frac{dT}{dP} = \frac{2(T_{tc} - T_0)}{P_{tc}} - 2\frac{(T_{tc} - T_0)}{P_{tc}^2} P$. Substituting this into (4) and using (15) the volume change is

$$\Delta V = \frac{9}{64} \theta \frac{(P_{tc} - P)^2}{P_{tc}^3}. \tag{16}$$

Alternatively, this expression can be found from (3) using (9) and (11). At $P = 0$, $\Delta V = \frac{9}{64} \theta \frac{1}{P_{tc}}$ and $\Delta V$ diminishes quadratically to zero as $P$ approaches $P_{tc}$.

The latent heat $L$ is obtained from (1) by inserting $T$ from (13) and $\Delta S$ from (15):

$$L = \frac{9\theta T_0}{128(T_{tc} - T_0)} + \frac{9\theta(2T_{tc} - 3T_0)}{128(T_{tc} - T_0)P_{tc}} P - \frac{27\theta}{128P_{tc}^2} P^2 + \frac{9\theta}{128P_{tc}^3} P^3. \tag{17}$$

$L$ is cubic in $P$ and has a maximum at $P = P_{tc}\left[1 - \sqrt{1 - \frac{2T_{tc}/3 - T_0}{T_{tc} - T_0}}\right]$. As $P$ approaches $P_{tc}$, $L$ decreases towards zero with a slope of $\frac{-9\theta T_{tc}}{128(T_{tc} - T_0)P_{tc}}$. At $P = 0$, $L = \frac{9\theta T_0}{128(T_{tc} - T_0)}$.

## 3. Melting curve for iron

Determining melting temperatures at high pressures has been a difficult task. The problem has persisted even though there have been great improvements in both experimental and numerical methods and the research community has been very active. The experiments at high pressures and temperatures are very demanding and the results are hard to obtain directly and accurately. Iron has achieved the most attention due to its importance in the Earth's core where it is believed to solidify from iron-rich melt to form the solid inner core: the liquid-solid interface is now at a pressure of 329 GPa [13].

The iron melting curve as developed during the last 20 years is presented in figure 3 and may be summarized as follows. A shock wave velocity investigation by Brown and McQueen [14] at very high pressures led to a calculated estimate at 243 GPa for the iron melting temperature (solid square). Williams *et al.* [15] were the first, using a laser heated diamond anvil cell (DAC), to conduct systematic static measurements of the iron melting temperature up to a pressure of about 100 GPa. The scatter (not shown here) in their data was several hundred K. They inferred the continuity of the melting curve (most tightly dotted line) at even higher pressures, up to about 250 GPa by also measuring, using a pyrometer, Hugoniot temperatures under shock conditions. Later Yoo *et al.* [16] also used a pyrometer for Hugoniot temperature measurements in their shock wave studies at the very high pressures of 235 and 300 GPa. They obtained melting temperatures (lower right corner triangles) again higher than Brown and McQueen.

Concurrently, Boehler [17] published static studies using a DAC up to 200 GPa, with considerably lower temperatures (widely dotted line). The scatter in his data points (not shown here) was a few hundred K. Saxena *et al.* [18] obtained slightly higher temperatures in their DAC experiments up to 150 GPa (short dashed line), with a scatter (not shown here) of a few hundred K. Jephcoat and Besedin [19] found, using also a laser heated DAC, an iron melting temperature at $P = 47$ GPa (open inverted triangle) with a result lower than Williams *et al.* but higher than the others.

Shen *et al.* [20] employed x-ray diffraction using synchrotron radiation to certify the presence of liquid or solid phase in a double-sided laser heated DAC. They reported temperatures (medium dashed line) with much reduced scatter and improved accuracy (typically $\pm$ 100 K) and identified the liquid-gamma-epsilon triple point to be at $P = 60$ GPa. Up to 60 GPa, their results were close to those measured by Boehler and Saxena *et al.* but with a steeper growth beyond that. At pressures 68 and 75 GPa they measured the highest temperatures where solid crystalline phase was observed and considered these to be a lower bound on the melting curve. In contrast, at $P = 23$ GPa they were able to measure with even higher accuracy the disappearance of gamma–iron diffraction peaks above $(2378\pm50)$ K and the reappearance of them when the temperature was lowered to $(2220\pm50)$ K (open square) demonstrating a small experimental hysteresis in this melting/solidification transition.

Iron melting has also been addressed by numerical simulations by three independent groups. The result of Laio *et al.* [21] (long dash-dotted line) lies partly between the results of Boehler and Saxena *et al.*, while the result of Belonoshko *et al.* [22] (short dash-dotted line) is between the results of Brown and McQueen, and Yoo *et al.* Alfè *et al.* have results from their ab initio calculations [23] (long dashed line) together with a free-energy correction (dashed-triple-dotted line) which improves the agreement with the experiments, thus being able to approach the result of Brown and McQueen and Shen *et al.*'s melting curve above 60 GPa. Later Belonoshko *et al.* [24] found bcc iron (for Alfè *et al.*'s model) has at $P = 323.5$ GPa a lower melting temperature (open cross) than their previously calculated epsilon iron melting temperature but higher than the calculations by Alfè *et al.* for Alfè *et al.*'s model of epsilon iron.

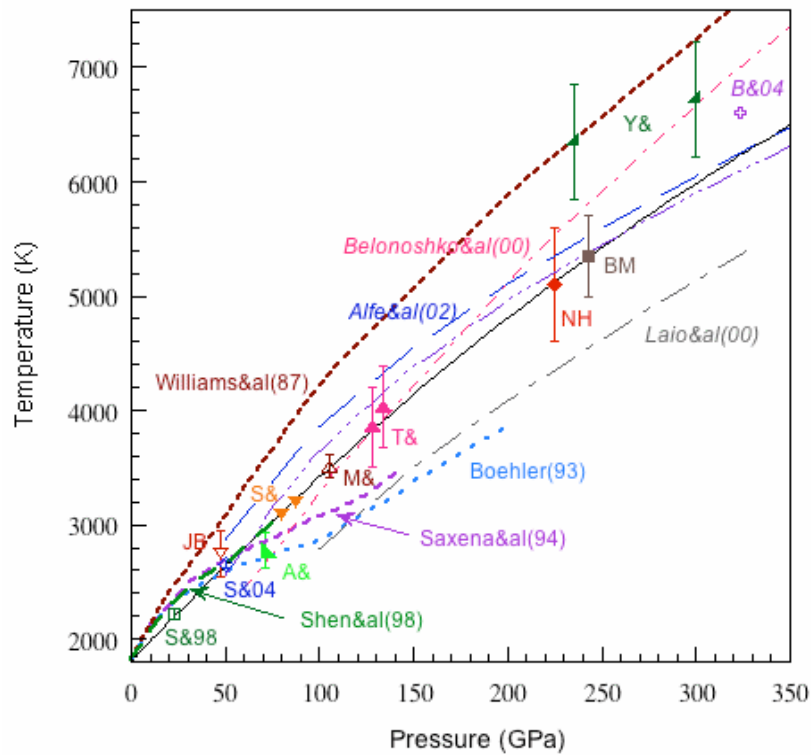

**Figure 3.** Recent development of the iron melting curve at high pressures. To help to identify the sources of the various curves, the first author and the two last digits of the year of publication are located near the curves. For the various symbols, Y& denotes Yoo et al., etc, BM denotes Brown and McQueen, etc, with the year shown where necessary. Italic font is used for the numerical results. The tightly dotted bold line shows static measurements up to 100 GPa joined to a high pressure shock wave temperature estimate [15]. Other static measurements are shown either by bold broken lines: widely dotted [17], short dashed [18] and medium dashed [20]; or open symbols: open square (solid appears) [20], inverted triangle [19], diamond (lowest melt) [27] and triangle [28]. Other shock wave results are shown by filled symbols: square [14], lower right corner triangles [16], lower left corner triangle [25], diamond [26], triangles [29] and inverted triangles [30]. Numerical simulations for epsilon iron are shown by long [21] or short [22] dash-dotted lines, and very long dashed and dashed–triple dotted lines [23], and for bcc iron by the open cross symbol [24]. Error bars are from the original sources. The solid line is the result of this theoretical work and is redrawn in figure 4.

A new shock wave study was published by Ahrens *et al.* [25] at $P = 71$ GPa: their calculated temperature (lower left corner triangle) was below Shen *et al.*'s melting curve. This result demonstrates that shock wave studies do not necessarily lead to higher temperatures than static measurements. By using an improved technique of shock waves, Nguyen and Holmes [26] have obtained a melting temperature (solid diamond) close to that of Brown and McQueen when a similar calculation was employed. Although both results have generous error bars the temperature and pressure values themselves indicate the iron melting curve is still increasing, not flattening out at those high pressures, and with a slope somewhat steeper than that obtained by Alfè *et al.* in their simulations.

Further DAC results using synchrotron radiation have appeared from Shen *et al.* [27] whose most accurately-bounded melting curve is at $P$ = 50 GPa: using scattering patterns they interpreted that the highest temperature solid occurred at $T$ = (2540 ± 55) K and the lowest temperature liquid at $T$ = (2650 ± 35) K (open diamond). Later Ma *et al.* [28] (open triangle) extended rather accurately (±100 K) the iron melting curve to the high pressure of 105 GPa where they observed the loss of all diffraction peaks at $T$ = 3510 K which "reflects the lower bound on the melting point". Their temperature is about 200 K below Alfè *et al.*'s simulation.

Two very recent shock wave studies by Tan *et al.* [29] (solid triangles) and Sun *et al.* [30] (inverted solid triangles) also demonstrate that there can be a smooth continuity from the accurate static results to the high pressure shock wave results of Nguyen and Holmes, and Brown and McQueen.

Since the experimental and numerical data have a rather large scatter a very careful consideration is here employed to choose the most reliable data available to facilitate the further analysis. The equation (13) gives the condition where the liquid phase becomes unstable as the order parameter ceases to be zero. This instability is not specific to the crystal structure of the appearing solid phase. According to the theory, for a pure substance the solidification temperature equals the melting temperature. Experimentally some hysteresis has been observed [20] for small $P$. Also, the experimental melting curves for small $P$ show some local curvature which this theoretical curve does not attempt to follow. Instead, for small $P$ we have two very accurate measurements for the instability of the liquid phase (at $P$ = 23 GPa [20] and $P$ = 50 GPa [27]) and only those are used in the analysis. Altogether, the following most trustworthy data points, shown in figure 4, have been selected covering as wide a pressure range as possible. First, the zero pressure melting temperature for iron is well-known: $T_0$ = 1811 K [31]. To find the coefficients for the linear and quadratic terms only the five most accurately measured DAC data points (backed-up by diffraction pattern analysis), were used, together with six agreeing shock-wave results at intermediate and high pressures where accurate static data were not available. The chosen DAC points are: the appearance of solid phase at 23 GPa, (2220 ± 50) K (the lowest open square) [20], the lowest temperature for the presence of only liquid phase closest to the melting curve which is found at 50 GPa, (2650 ± 35) K (open diamond) [27], and the three slightly less accurate x-ray diffraction measured highest temperatures for crystalline epsilon-iron: the points at 68 and 75 GPa (two higher open squares, with a typical error bar of ± 100 K, from fig. 3 in [20]) and the point at 105 GPa with an accuracy of ±100 K (open triangle) [28]. All shock wave results are much less accurate and from them were chosen all consistent data points: the very recent results [30] (inverted solid triangles) and [29] (solid triangles) as well as the well-established higher pressure results [26] (solid diamond) and [14] (solid square).

Figure 4 presents $T_0$ and these five most accurate DAC experimental points and the six consistent shock wave results together with the best fit curve to the equation (13). These twelve data points differ from the curve by less than 3 %, ten of them by less than 1.2 %. The solidification curve flattens at very high pressures and ends at the tricritical point $(P_{tc}, T_{tc})$ = (793 GPa, 8632 K) whose lower estimate is (682 GPa, 7800

K) and higher estimate is (904 GPa, 9465 K). The result for $T_{tc}$ agrees well with the literature range of $T_c$, 5970 – 10000 K [32]. The initial slope of the melting curve at zero pressure is $dT/dP = 2(T_{tc} - T_0)/P_{tc}$ giving 17.2 K/GPa. This is about half of the estimate quoted in the literature [33, 34]. This is due to the quadratic growth of this solidification curve from $T_0$ via the first accurate data point at 23 GPa where melt solidifies while the earlier estimates are based on older measurements on melting of solid at smaller pressures, and those results show a different local curvature at small $P$, not possible for this model covering a wide pressure range.

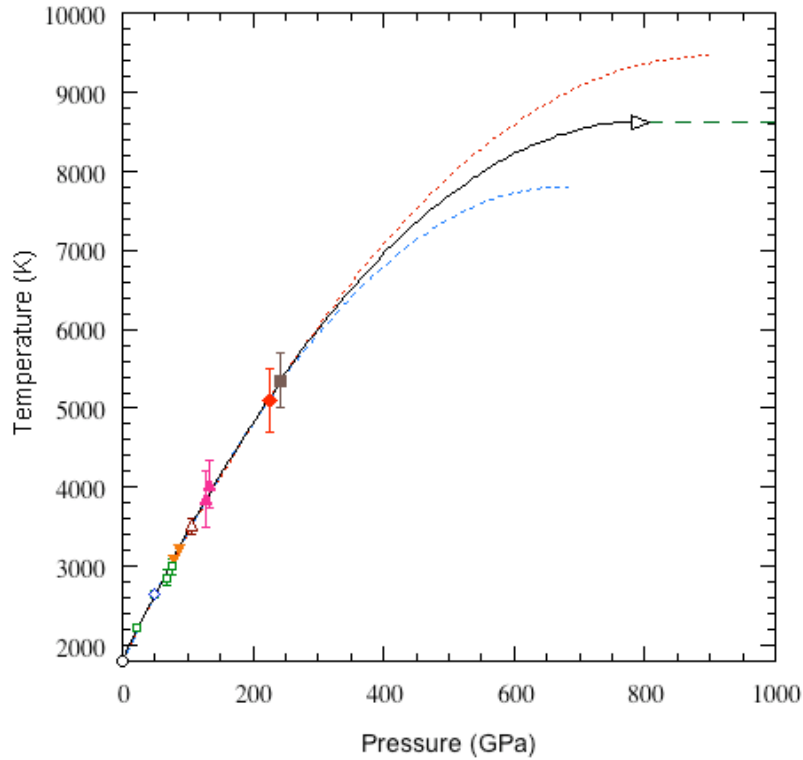

**Figure 4.** Solidification temperature $T$ of the iron melt as a function of pressure $P$ as calculated from equation (13) (solid line) with its maximum uncertainty (dashed lines) and the consistent experimental data. The most accurate experimental results used to find the curvature are shown as open symbols: circle for $T_0$, three squares [20], diamond [27] and triangle [28]. All the supporting shock wave results are shown by solid symbols: square [14], diamond [26], triangles [29] and inverted triangles [30]. Error bars are from the original sources, but [30] gave none. The triangle pointing right is for the tricritical point. The horizontal dashed line is for $T = T_{tc}$.

This theoretical melting curve lies rather well in the middle of the large range of all the experimental results and agrees within a few hundred K with the ab initio calculations. By using the maximum and minimum estimates for the tricritical point one can bound the precision of the pure iron solidification temperature within 0.3 – 2 % at three geophysically interesting pressures [13]: At the core–mantle boundary (CMB) $P_{CMB} = 136$ GPa gives $T_{CMB} = (3945 \pm 12)$ K, at the inner core boundary (ICB) $P_{ICB} = 329$ GPa gives $T_{ICB} = (6290 \pm 80)$ K and at the centre of the Earth $P_{centre} = 364$ GPa gives $T_{centre} = (6630 \pm 110)$ K. The first of these is slightly lower than what is obtained from Alfè *et al.*'s ab initio calculations, the second slightly higher and the third a few hundred K higher, but they all have much better precision than has been obtained by any other methods.

The upper part of the melting curve and its extension beyond the tricritical point give an estimate for the phase transition line in the so-called warm dense matter (WDM) regime [35], which is the high energy density regime between plasma physics and condensed matter. WDM is important, for instance, in confined fusion and in the interiors of the large planets. In WDM, strongly-coupled plasma exhibits long- and short-range order allowing, if cooled, a phase transition from plasma to a solid state. These phenomena are presently very challenging both theoretically and experimentally. It is worth considering that there is a tricritical point on the threshold between dense cool plasma and condensed matter joining the transitions from ionized gas to liquid and to solid. The iron tricritical pressure is at present slightly beyond what is reached by the recent high power laser driven shock techniques [36] used to study WDM; through these techniques, iron Hugoniot temperatures have been measured at pressures up to 700 GPa, unfortunately still with rather substantial error bars.

## 4. Latent heat and the change of entropy and volume for iron as function of pressure

All the thermodynamic quantities need some experimental calibration to produce results in physical units. Only at zero pressure is the latent heat of iron well known: 247.3 kJ kg$^{-1}$ [31] or 13.81 kJ mol$^{-1}$. This is for melting and solidification of bcc delta phase. The crystal structure and the magnetic properties of the solid phase can change along the melting curve as pressure increases: the bcc delta phase is followed first by the fcc gamma phase, both paramagnetic, and then by the nonmagnetic hcp epsilon phase and possibly by the bcc phase at even higher pressures. The possible effects of these changes on the calculated values are not taken into account by this theory. The calibration here is solely based on the known value of $L$ at zero pressure due to the lack of accurate data at higher pressures. By using the known value of $L$ with the value of $T_0$, the change in entropy at zero pressure can be calculated from (1) to be 7.626 J mol$^{-1}$ K$^{-1}$. Together with the slope of the melting curve at $P = 0$, this gives $\Delta V = 0.1311$ cm$^3$ mol$^{-1}$ from (4). This is about half the value in the literature (see, for instance, [33, 34]), as needs to follow from the slope difference discussed earlier.

By comparing the zero pressure value for $\Delta S$ with (15) one can solve for the dimensional multiplier of the Landau potential for iron:
$$\theta = \frac{128 \times 7.626}{9}(T_{tc} - T_0) \, \text{J mol}^{-1}\text{K}^{-1} = 739.8 \, \text{kJ mol}^{-1}.$$

The pressure dependences of $\Delta V$ and $L$ are presented in figure 5. $\Delta V$ decreases monotonically but $L$ has a maximum at $P = 278$ GPa. At the tricritical point both $\Delta V$ and $L$ vanish as does $\Delta S$. Figure 6 shows $\Delta S$ as a function of $\Delta V$. The behaviour of $\Delta S$ as $\Delta V$ approaches zero differs from the result for gas condensation where $\Delta S$ approaches a finite value, $R \ln 2$, as $\Delta V$ vanishes. The linear ratio of $\Delta S$ and $\Delta V$ has been considered valid for some metals' melting, too, but those data are not close to the origin (see discussion in [33] and references therein). Here, for iron, $\Delta S = R \ln 2$ already at $\Delta V = 0.075$ cm$^3$ mol$^{-1}$ and above it the relationship between $\Delta S$ and $\Delta V$ can rather well be approximated by a line.

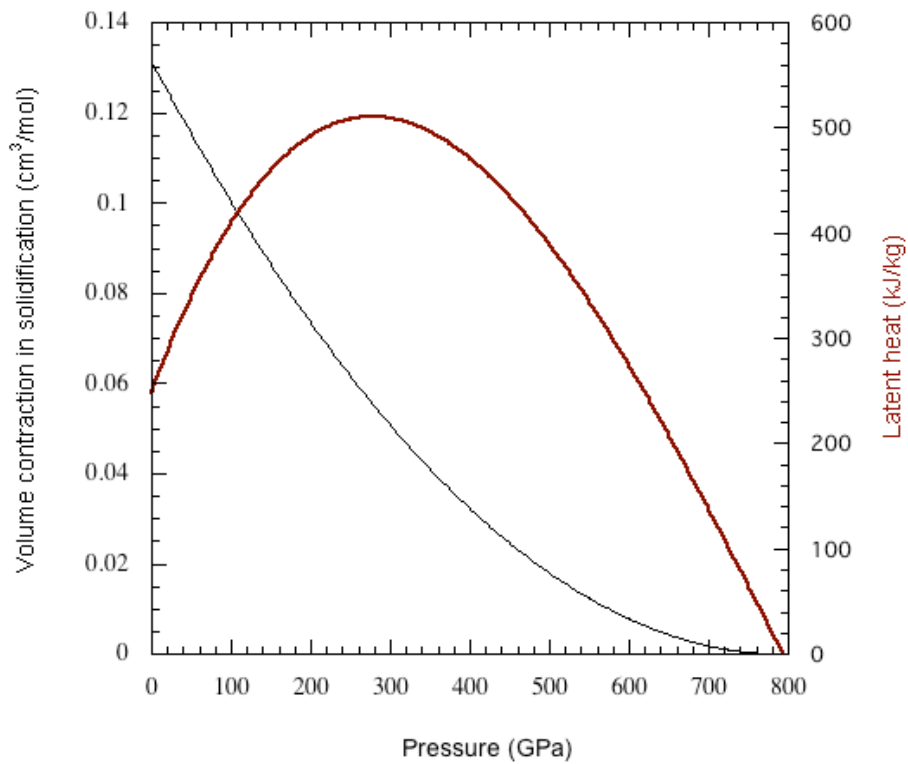

**Figure 5.** Pressure dependence of the volume contraction in the solidification together with the latent heat (bolder line).

$L$ can now be determined at the geophysically interesting core pressures where previously it has been possible to estimate it only rather unsatisfactorily [37], it being one of the most uncertain parameters in the Earth's heat budget calculations. It has been estimated to be as low as 400 kJ/kg [38], or as high as 1560 kJ/kg [39]. Here $L$ agrees with the low part of this range since its maximum value at (280±50) GPa is (510±40) kJ/kg. The uncertainties follow from the higher and lower estimates of the tricritical point. At the centre of the Earth one finds $L$ is $\left(490^{+60}_{-80}\right)$ kJ/kg, at the inner core boundary $\left(500^{+50}_{-60}\right)$ kJ/kg and at the core-mantle boundary $\left(447^{+10}_{-14}\right)$ kJ/kg. At the inner core boundary the value of $L$ is 10 % smaller than its estimate in [33] calculated using a temperature about 26 % less and melting entropy about 30 % more than those found here.

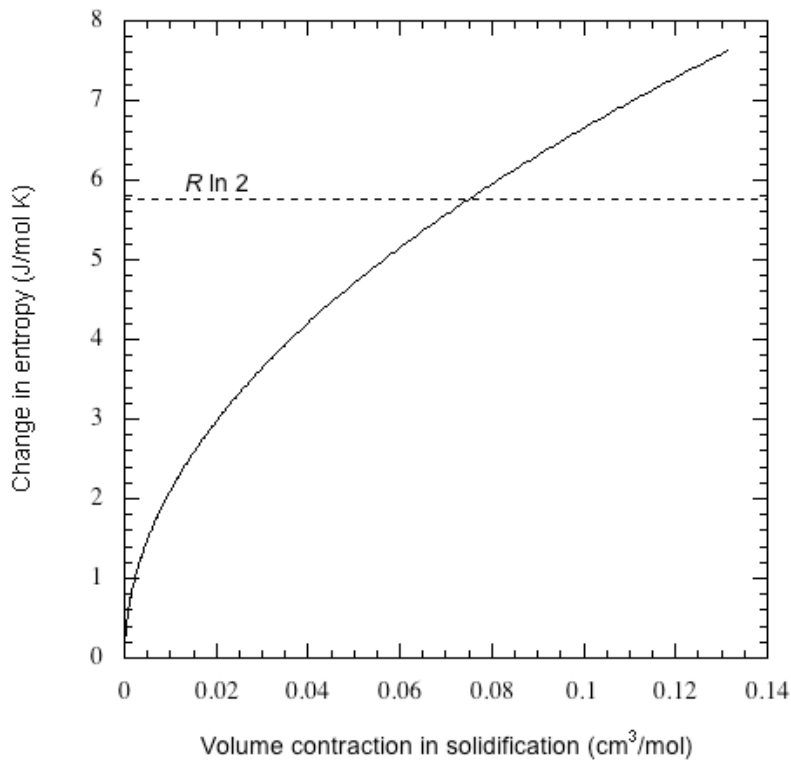

**Figure 6.** The change of entropy as function of the volume contraction in the solidification. Dashed line shows $\Delta S = R \ln 2$ which is reached at $\Delta V = 0.075$ cm$^3$/mol corresponding to a pressure of 194 GPa.

## 6. Conclusions

This work depends on the assumption that a solidification/melting curve can end at a tricritical point. Since this tricritical point has remained beyond experimental techniques for three-dimensional matter, the grounding of this assumption is based on theoretical expectations together with the available phase transition data and with the supporting evidence of a tricritical point at the end of the melting curve in examples of two-dimensional matter. Landau theory allows one to find the general expression for the solidification or melting curve if it ends at a tricritical point where $dT/dP = 0$. This gives a quadratic pressure dependence for the solidification temperature, with a definite structure of the coefficients. For pure iron this temperature formula agrees very well with twelve experimental results in a pressure range of 0 – 250 GPa. Thus it is likely that iron has a tricritical point at the very high pressure of $(800 \pm 100)$ GPa. It follows from the quadratic pressure dependence that one can find the iron solidification temperatures which occur at the high pressures of the Earth's core. These are upper bounds on the true temperatures there, since there are also impurities like nickel with some light elements which reduce the solidification temperature from that of pure iron. In addition, one obtains for the very first time systematic estimates for latent heat, volume change and entropy change in the solidification as a function of pressure.